\documentclass[f]{ceurart}
\usepackage[utf8]{inputenc}
\usepackage{listings}
\usepackage{enumitem}
\usepackage{xcolor}

\begin{document}
\copyrightyear{2024}
\copyrightclause{Copyright for this paper by its authors.
  Use permitted under Creative Commons License Attribution 4.0
  International (CC BY 4.0).}
\conference{Under review for CLAIRvoyant - ConventicLE on Artificial Intelligence Regulation - 2024 @ Jurix 2024}
\title{Developing an Ontology for AI Act Fundamental Rights Impact Assessments}
\author{Tytti Rintamäki}
\cormark[1]
\fnmark[1]
\author{Harshvardhan J. Pandit}
\cormark[1]
\fnmark[1]
\address{ADAPT Centre, School of Computing, Dublin City University, Dublin, Ireland}
\cortext[1]{Corresponding author: Tytti Rintamäki tytti.rintamaki@adaptcentre.ie and Harshvardhan J. Pandit me@harshp.com}
\fntext[1]{These authors contributed equally.}

\begin{abstract}
The recently published EU Artificial Intelligence Act (AI Act) is a landmark regulation that regulates the use of AI technologies. One of its novel requirements is the obligation to conduct a Fundamental Rights Impact Assessment (FRIA), where organisations in the role of deployers must assess the risks of their AI system regarding health, safety, and fundamental rights. Another novelty in the AI Act is the requirement to create a questionnaire and an automated tool to support organisations in their FRIA obligations. Such automated tools will require a machine-readable form of information involved within the FRIA process, and additionally also require machine-readable  documentation to enable further compliance tools to be created. In this article, we present our novel representation of the FRIA as an ontology based on semantic web standards. Our work builds upon the existing state of the art, notably the Data Privacy Vocabulary (DPV), where similar works have been established to create tools for GDPR's Data Protection Impact Assessments (DPIA) and other obligations. Through our ontology, we enable the creation and management of FRIA, and the use of automated tool in its various steps.
\end{abstract}
\begin{keywords}
\sep AI Act \sep Risk Assessment \sep FRIA \sep Fundamental Rights \sep Ontology \sep Semantic Web
\end{keywords}
\textcolor{red}{Presented at CLAIRvoyant (ConventicLE on Artificial Intelligence Regulation) Workshop 2024}
\maketitle

\section{Introduction}
The European Union's recently published Artificial Intelligence Act (AI Act) \cite{EU_AIAct} is the first of its kind regulation that governs AI systems with a particular focus on harms to health, safety, and fundamental rights. A key and novel requirement established in AI Act Article 27 is the Fundamental Rights Impact Assessment (FRIA) which requires deployers of AI systems to assess the risks and impacts of their AI systems on fundamental human rights. The FRIA is a structured process following existing procedures for impact assessments, and was developed based on the similar procedure under the General Data Protection Regulation (GDPR) \cite{EU_GDPR} for Data Protection Impact Assessments (DPIA). As development of AI technologies has progressed rapidly within the last decade, and the AI Act itself is a new legal framework, conducting and using a FRIA poses significant challenges not only regarding legal compliance, but also from the perspectives of data governance to identify and maintain relevant information and information systems to develop tools that can support and enhance these processes.

The current conventional method for implementing FRIA is to identify obligations linked to specific clauses in the AI Act and find the steps needed to complete them. For organisations, such tasks are primarily human-oriented activities that utilise word processor software (e.g. MS Word) and document formats (e.g. PDF) that contain unstructured information and are not suitable for developing automated procedures and tooling. Further, organisations commonly have several departments or organisational units that can contain different technologies and practices, making a combined legal assessment of the organisation as a whole a challenging and complicated affair. The AI Act in Article 27-5 foresees such challenges and requires the AI Office, the EU body responsible for the implementation of the AI Act, to create a `questionnaire' to support organisations in meeting the obligations for a FRIA. The AI Act, in the same article, also states that such a questionnaire should be provided with an automated tool - though it does not clarify what automation or support mean in this context.

Based on prior work regarding the use of knowledge engineering and information systems to support and enhance the DPIA process and its use for complying with the GDPR \cite{panditSemanticSpecificationData2022}, we identify the need for a formal representation of the FRIA process that can aid the creation of questionnaires and automated tools and support the documentation involved by providing a consistent, structured, and interoperable approach. By developing an ontology for FRIA, this paper thus aims to bridge the gap between the high-level legal requirements of the AI Act and the technical, procedural steps necessary for effective compliance, risk management, and governance of AI systems.

We use the following research question to guide our work: ``\textit{How can we represent the FRIA as an organisational process through a standards-based, machine-readable, and interoperable ontology?}''. Here, standards refers to the W3C semantic web standards of RDF for representing information in a machine-readable format, and RDFS and OWL to create an ontological representation. The contribution of this article is thus an OWL ontology that enables the operational implementation of a FRIA within an organisation to meet the AI Act's obligations.

\section{Rationale}
To develop the ontology, we follow the Linked Open Terms (LOT) ontology engineering methodology \cite{poveda2022lot} which has been used successfully in several projects, and is based on the NeOn ontology engineering methodology which has been used in the creation of similar legally relevant ontologies including for GDPR's DPIA \cite{panditSemanticSpecificationData2022}. The first step in LOT is the development of an ontology requirements specification that outlines ``why the ontology is being built and to identify and define the requirements the ontology should fulfil''. For this, LOT recommends the use of competency questions which are a well established practice in the ontology engineering community. We reused the template provided by LOT to generate the ontology requirements specification, which is presented in Table 1.

\begin{table*}[ht]
\centering
\caption{FRIA Ontology}
\label{tab:ORSD}
\scriptsize
\begin{tabular}{| l | l | l | l  | l | l | l |l| }
\hline
\multicolumn{8}{|c|}{\cellcolor[HTML]{A0A0A0}\textbf{}} \\ \hline
\multicolumn{8}{|c|}{\cellcolor[HTML]{EFEFEF}\textbf{1. Purpose}} \\ \hline
\multicolumn{8}{| p{14.5cm} |}{The purpose of this ontology is to model the FRIA as an information process.} \\ \hline
\multicolumn{8}{|c|}{\cellcolor[HTML]{EFEFEF}\textbf{2. Scope}} \\ \hline
\multicolumn{8}{| p{12.0cm} |}{The scope of this ontology is limited to modelling the FRIA as defined in AI Act Article 27.} \\ \hline
\multicolumn{8}{|c|}{\cellcolor[HTML]{EFEFEF}\textbf{3. Implementation Language}} \\ \hline
\multicolumn{8}{| p{12.0cm} |}{W3C semantic web standards - OWL, RDFS, SKOS} \\ \hline
\multicolumn{8}{|c|}{\cellcolor[HTML]{EFEFEF}\textbf{4. Intended End-Users}} \\ \hline
\multicolumn{8}{| p{12.0cm} |}{Organisations who create and use FRIA i.e. deployers of the AI systems and AI Act authorities} \\ \hline
\multicolumn{8}{|c|}{\cellcolor[HTML]{EFEFEF}\textbf{5. Intended Uses}} \\ \hline
\multicolumn{8}{| p{12.0cm} |}{
Use 1. To document obligations regarding FRIA.\newline
Use 2. To document outcomes of FRIA.\newline
Use 3. To notify authorities regarding FRIA.
 } \\ \hline
\multicolumn{8}{|c|}{\cellcolor[HTML]{EFEFEF}\textbf{6. Ontology Requirements}} \\ \hline
\multicolumn{8}{|c|}{\cellcolor[HTML]{EFEFEF}\textbf{a. Non-Functional Requirements}}    \\ \hline
\multicolumn{8}{| p{12.0cm} |}{
NFR 1. Interoperability: The ontology must extend existing legal compliance ontologies (e.g. DPV) \newline
NFR 2. Scalability: The ontology should be adaptable/extensible for future developments.\newline
NFR 3. Usability: The ontology should support use by legal and non-legal stakeholders.
} \\ \hline
\multicolumn{8}{|c|}{\cellcolor[HTML]{EFEFEF}\textbf{b. Functional  Requirements: Groups of Competency Questions}}  \\ \hline
\multicolumn{4}{|c|}{\cellcolor[HTML]{EFEFEF}CQG1. Related to AI Act obligations} & \multicolumn{4}{|c|}{\cellcolor[HTML]{EFEFEF}CQG2. Related to Organisational Governance} \\ \hline 
\multicolumn{4}{ | m{7.25cm} |}{
CQ1. When was the FRIA conducted? \newline 
CQ2. What is the Intended Purpose of the AI system? \newline 
CQ3. What are the risks, consequences, impacts? \newline 
CQ4. What are the mitigation measures?
}        & 
\multicolumn{4}{ m{7.25cm} |}{
CQ5. What is the outcome of the FRIA process? \newline 
CQ6. What fundamental rights are affected? \newline 
CQ7. What authorities are notified for the FRIA?  \newline
CQ8. What documentation/tools were used for FRIA?
}\\ \hline

\end{tabular}
\vspace{-0.1in}
\end{table*}

\section{State of the Art}
Within ontology engineering methodologies based on semantic web, including LOT, the reuse of existing ontologies is heavily recommended. Therefore, following the requirements specification, we explore the state of the art to identify relevant resources and assess the extent to which they can be reused to implement our FRIA ontology. In this, we divide the existing literature in two categories: first, existing ontologies that directly and explicitly address FRIA, or if not, then the AI Act; and second, ontologies that address similar mechanisms such as DPIA under GDPR, or impact assessments based on established procedures such as ISO standards.

\subsection{Existing Ontologies for FRIA and AI Act}
Given the recency of the AI Act in terms of development and finalisation, few approaches have emerged that provide ontologies modelling it. Golpayegani et al. were one of the first approaches to model the requirements of the AI Act as an ontology through the AI Risk Ontology (AIRO) \cite{golpayegani2022airo} - an OWL2 ontology based on an early draft version of the AI Act. AIRO provides a risk management ontology based on the requirements of the AI Act and ISO standards, and acts as the upper ontology for Vocabulary of AI Risks (VAIR) \cite{golpayegani2023high}. Golpayegani et al. have demonstrated the use of AIRO and VAIR to model the high-risk use-cases defined in AI Act Annex III as a logical group of semantic concepts and showed that logical reasoning or validation approaches such as SHACL can be used to determine whether the use-case is high-risk under the AI Act \cite{golpayegani2023high}. Golpayegani et al. have also developed the AI Card \cite{golpayeganiAICardsApplied2024} as a visual approach for documenting the AI system through the use of AIRO and VAIR as its machine-readable representation. AIRO and VAIR provide concepts required in the AI Act such as stakeholders, AI processes, and risk management, but do not contain a modelling of the FRIA.

Hernandez et al. developed the Trustworthy AI Requirements Ontology (TAIR)  \cite{hernandez2024open} which models the clauses of the AI Act and of relevant ISO standards as a series of requirements and compares them to identify where such standards would be useful for compliance with the AI Act. This work addresses the requirement for using `Harmonised Standards' in the AI Act, and includes concepts regarding impact assessments and risk management, albeit these are based on a draft version of the AI Act.

Though not providing an ontology or directly addressing the final AI Act requirements, these works provide an exploration of the FRIA requirements by identifying information involved in conducting a FRIA: the FRIA template produced in the ALIGNER H2020 project \cite{FundamentalRightsImpact}, an analysis of the FRIA requirements in AI Act by Mantelero \cite{manteleroFundamentalRightsImpact2024},  the rights impact assessments for algorithmic systems by Gerards et. al \cite{gerardsFundamentalRightsAlgorithms2022}, the algorithmic impact assessment published by the Govt. of Canada \cite{governmentofcanadaAlgorithmicImpactAssessment2024}, a quantified risk score for impact on fundamental rights by Inverardi et. al. \cite{inverardiFundamentalRightsAI2024}, a method for assessing the severity of impacts on fundamental rights by Malgieri and Santos \cite{malgieriAssessingSeverityImpacts2024}, and an interpretation of the draft AI Act's FRIA requirements by Janssen et. al \cite{janssenPracticalFundamentalRights2022}

\subsection{Existing Ontologies for DPIA, GDPR, and Impact Assessments}

In comparison to the AI Act, the GDPR has been in effect for 6 years, and has been addressed through several surveys on compliance approaches and developed ontologies \cite{esteves2024analysis,kurteva2024consent,zaguir2024challenges}. Notable approaches in these include the ontology of privacy requirements by Gharib et al. \cite{gharib2020ontology}, The core ontology for privacy requirements (CoPri), \cite{gharib2021copri} which provides a framework for modelling legal processes and requirements, Privacy Ontology for Legal Reasoning (PrOnto) \cite{palmirani2018pronto} which provides concepts with the aim of modelling legal norms and assessing them through deontic reasoning, and the Data Privacy Vocabulary (DPV) \cite{panditDataPrivacyVocabulary2024} which is an output of the W3C Data Privacy Vocabularies and Controls Community Group (DPVCG), and provides an extensible collection of vocabularies for modelling legal concepts associated with data and technologies.

Of these, only the DPV has the community and infrastructure supporting the continuos development and refinement of the resource, and is also the only resource that had modelled GDPR which has been expanded to also model the AI Act using the same core concepts \cite{panditDataPrivacyVocabulary2024}. The DPV is also the only resource we know of that provides rich taxonomies to represent real-world concepts associated with the ontological concepts e.g. for purposes and data categories. The DPV features a TECH extension which provides concepts for modelling the technology lifecycle, stakeholders, and documentation, the RISK extension for modelling risk assessments, and the AI extension which extends these to provide AI-specific concepts. In DPV, the legal concepts derived from specific regulations are provided in a separate namespace from these other `core' vocabularies, and DPV provides such legal extensions for EU GDPR and the EU AI Act. The DPV's GDPR extension provides concepts modelling the DPIA process based on \cite{panditSemanticSpecificationData2022}. At the moment, the DPVCG is integrating AIRO \cite{golpayegani2022airo} and VAIR \cite{golpayegani2023high} in to the AI and AI Act extensions.

\section{FRIA Ontology}
Our objective is to develop an ontology to model the FRIA as defined in the AI Act Article 27 as an information process through which stakeholders such as deployers and authorities can create automated technological tools to support the compliance activities. For ensuring our ontology is interoperable and extensible, we utilise semantic web standards such as RDF to represent it, SKOS to create a vocabulary or thesauri, and RDFS and OWL2 for knowledge representation. We follow the Linked Open Terms (LOT) \cite{poveda2022lot} as the methodology for ontology engineering, which strongly recommends reusing existing ontologies where relevant. For this, from Section 3, we identified AIRO \cite{golpayegani2022airo} and VAIR \cite{golpayegani2023high} as the most relevant ontologies for the AI Act, and the DPV \cite{panditDataPrivacyVocabulary2024} as a useful resource for practical use of legal concepts. Since AIRO and VAIR are being integrated for the upcoming DPV version 2.1, we aim to support this integration by identifying the concepts not present in these existing ontologies.

DPV is provided with RDFS+SKOS semantics as the `default serialisation', with a separate namespace used for OWL2 semantics to support the use of concepts beyond the strict requirements of logical constraints when using OWL2. Following this, we provide the concepts necessary to model the FRIA in this article which can then be expanded as more information is available to represent real-world constraints and logical assertions using OWL2 or another method.

The concept for Fundamental Rights Impact Assessments (FRIA) already exists within the main DPV as \texttt{dpv:FRIA}, and is extended as \texttt{eu-aiact:FRIA} in AI Act extension to represent the FRIA as defined within the AI Act. This concept represents the FRIA as both an activity and as an artefact (e.g. a document). Therefore, in our FRIA ontology, we create separate explicit concepts for modelling the information and steps involved in the DPIA process by expanding this central concept.

Based on the interpretation of the FRIA in AI Act Article 27, and by using existing work interpreting the DPIA in GDPR as a series of steps that are represented through an ontology \cite{panditSemanticSpecificationData2022}, we identified the following groups of concepts for our ontology:
\setlist{nolistsep}
\begin{enumerate}
  \item \textbf{FRIA Metadata:} concepts representing relevant metadata regarding the FRIA such as when it was conducted, by whom, for which AI systems, etc.;
  \item \textbf{FRIA Necessity:} concepts representing the step where a necessity for conducting a FRIA is identified as per AI Act Article 27-1;
  \item \textbf{FRIA Inputs:} concepts representing the inputs required in a FRIA as per AI Act Article 27-1, and the reuse of a DPIA as per AI Act Article 27-4;
  \item \textbf{FRIA Outcomes:} concepts representing the outcomes identified from conducting a FRIA as per AI Act Article 27-1;
  \item \textbf{FRIA Notifications:} concepts representing the step where a FRIA to be communicated to an authority as per AI Act Article 27-3;
  \item \textbf{FRIA Automated Tools:} the use of questionnaire and/or automated tools as per AI Act Article 27-5.
\end{enumerate}

We use the following namespaces and prefixes in describing our proposed ontology:
\setlist{nolistsep}
\begin{itemize}
  \item Our proposed ontology: \textit{https://example.com/FRIA\#} with prefix \texttt{fria:}. 
  \item DCMI Metadata Terms: \texttt{http://purl.org/dc/terms/} with prefix \texttt{dct}.
  \item DPV: \textit{https://w3id.org/dpv\#} with prefix \texttt{dpv:}.
  \item DPV TECH extension: \textit{https://w3id.org/dpv/tech\#} with prefix \texttt{tech:}.
  \item DPV RISK extension: \textit{https://w3id.org/dpv/risk\#} with prefix \texttt{risk:}.
  \item DPV AI extension: \textit{https://w3id.org/dpv/ai\#} with prefix \texttt{ai:}.
  \item DPV EU AI Act extension: \textit{https://w3id.org/dpv/legal/eu/aiact\#} with prefix \texttt{eu-aiact:}.
\end{itemize}

\subsection{Metadata for FRIA}
A FRIA, as a documentation requirement under the AI Act, is expected to be regularly updated as per Article 27-2 ``the deployer shall take the necessary steps to update the information''. Therefore it is necessary to indicate temporal information and provenance associated with the FRIA. For these, we reuse prior work establishing the reuse of DCMI terms for GDPR's DPIA \cite{panditSemanticSpecificationData2022} regarding temporal information (\texttt{dct:created}, \texttt{dct:modified}, \texttt{dct:dateSubmitted}, \texttt{dct:dateAccepted}, \texttt{dct:temporal}, \texttt{dct:valid}), conformance e.g. codes of conduct (\texttt{dct:conformsTo}), descriptions (\texttt{dct:title}, \texttt{dct:description}), identifier or version (\texttt{dct:identifier}, \texttt{dct:isVersionOf}), and subject or scope (\texttt{dct:subject}, and \texttt{dct:coverage}).

To record provenance, we suggest reusing \texttt{dct:publisher} to record the organisation responsible for conducting the FRIA, \texttt{dct:contributor} to denote the personnel and entities involved, and \texttt{dct:provenance} to refere to a log of changes. Additionally, \texttt{dct:creator} can record the specific entity or tool used to `create' the resource - which is relevant as the AI Act Article 27-5 explicitly provides for the use of automated tools in the FRIA process.

\subsection{Concepts to represent necessity of FRIA}
AI Act Article 27-1 describes the conditions under which a FRIA is necessary. An organisation therefore has the obligation to first assess whether it must conduct a FRIA. We represent this process through the concept \texttt{fria:FRIANecessityAssessment} which extends the existing \texttt{eu-aiact:FRIA} concept and can be associated with a FRIA using the relation \texttt{dpv:hasAssessment}. To represent the specific outputs of this process, we create the concepts \texttt{fria:FRIANecessityStatus} which is associated with the assessment using the relation \texttt{dpv:hasStatus}. To represent specific outcomes, we create the instances \texttt{fria:FRIARequired} and \texttt{fria:FRIANotRequired}.

\subsection{Concepts to represent inputs of FRIA}
We represent the process of carrying out the FRIA as the concept \texttt{fria:FRIAProcedure} which extends the existing \texttt{eu-aiact:FRIA} concept and can be associated with a FRIA using the relation \texttt{dpv:hasAssessment}. AI Act Article 27-1 describes the information which must be included when conducting a FRIA, which we interpret as follows: 

\begin{enumerate}
  \item Article 27-1a description of the deployer’s processes: represented by extending \texttt{dpv:Process} as the concept \texttt{fria:AIProcess}, and associated using the relation \texttt{dpv:hasProcess}. This follows the DPV's modelling of similar processes for GDPR where a \texttt{dpv:Process} provides a way to combine other concepts such as purposes, data, technologies, and entities in specific roles.
  \item Article 27-1a intended purpose: represented by the existing \textit{eu-aiact:IntendedPurpose} whose parent is \textit{dpv:Purpose}, and is assocaited using the relation \texttt{dpv:hasPurpose}. Note that the AI Act's intended purpose is a broad concept that goes beyond DPV's modelling of purpose as referring to the objective or goal, whereas intended purpose includes details such as the AI technique and data involved. Therefore, we suggest also modelling \texttt{eu-aiact:IntendedPurpose} as the subclass of \texttt{dpv:Process}.
  \item Article 27-1b period of time: represented using \textit{dpv:Duration} and associated using the relation \texttt{dpv:hasDuration}. DPV provides enumerated concepts for different durations such as endless, fixed, temporal, until event, and until time.
  \item Article 27-1b frequency: represented using \textit{dpv:Frequency} and associated using the relation \texttt{dpv:hasFrequency}. DPV provides enumerated concepts for different frequencies such as continous, often, singular, and sporadic.
  \item Article 27-1b intended to be used: represented as \textit{fria:IntendedUse}, where we interpret this concept to be different from \textit{aiact:IntendedPurpose} as the `purpose' is a declaration of why the AI system is needed or to be used, and `use' is the contextual application of that purpose in specific scenarios or `deployments'. The same AI system with one intended purpose can thus have different intended uses in separate scenarios e.g. through variance in input and output data (including decisions produced), human subjects involved, hardware and software being used. DPV already contains the concept \texttt{tech:IntendedUse}, which should be the parent of this concept.
  \item Article 27-1c categories of natural persons and groups: which can be represented using the existing concept \texttt{dpv:DataSubject} provided in DPV with a taxonomy of categories such as tourists, adults, minors. Though, data subject in DPV is currently defined as per the GDPR in terms of natural persons whose data is being processed, which is not compatible with our intended concept here. AIRO has \texttt{AISubject} which is defined as natural persons subjected to the use of AI which is more in line with what we want to model. Therefore, to avoid duplication of these categories under AISubject, and to avoid backwards incompatible changes to DPV as it is being actively used, we propose either changing the definition of data subject to include `data and AI subjects' - following a similar change made in DPV 2.0 where the term remains the same \texttt{dpv:DataSubject} but its use now encompasses any data or technology including AI. Or, to create a new concept called \texttt{dpv:HumanSubject} as the parent of \texttt{dpv:DataSubject} and \texttt{airo:AISubject}, and move the taxonomy of data and AI subjects under it.
  \item Article 27-1c likely to be affected by its use in the specific context: where likely is represented by \texttt{dpv:Likelihood} and associated using \texttt{dpv:hasLikelihood}. Affected is interpreted as referring to a \texttt{dpv:Consequence} taking place with its subcategory \texttt{dpv:Impact} - which are associated using \texttt{dpv:hasConsequence} and \texttt{dpv:hasImpact} respectively. The entity or thing being affected is associated using \texttt{dpv:hasConsequenceOn} and \texttt{dpv:hasImpactOn} respectively. In DPV, consequence is the general term for referring to events such as failure of equipment, and impact is the preferred term for referring to entities being (significantly) affected such as through physical harm or loss of resources.
  \item Article 27-1d risks of harm: are represented by using \texttt{dpv:Risk} for the risk, with the concept \texttt{risk:Harm} to refer to harm. The RISK extension provides further concepts for modelling different categories of harms e.g. \texttt{risk:PhysicalHarm}. It also models these concepts as \texttt{dpv:} to enable their use in different roles across use-cases e.g. \texttt{risk:Harm} as a risk source, risk, consequence, or impact through the use of relevant relations.
  \item Article 27-1e human oversight measures: represented using \texttt{dpv:HumanInvolvementForOversight} along with other \texttt{dpv:HumanInvolvementConcepts}. DPV also provides additional taxonomies to represent whether the entity can perform some activity (\texttt{dpv:EntityPermissiveInvolvement}) such as correcting outputs or cannot perform an activity (\texttt{dpv:EntityNonPermissiveInvolvement}) such as not being able to opt out.
  \item Article 27-1e instructions for use: this is represented using \texttt{eu-aiact:InstructionsForUse} which extends from \texttt{tech:Documentation}.
  \item Article 27-1f measures to be taken in the case of the materialisation of those risks: this is represented using \texttt{dpv:RiskMitigationMeasure}, where the specific measures mentioned in this clause include - arrangements for internal governance and complaint mechanism, represented by \texttt{dpv:GovernanceProcedures} and its more specific form \texttt{dpv:IncidentManagementProcedures} and \texttt{dpv:IncidentReportingCommunication}. 
\end{enumerate} 

In addition to these, Article 27-2 mentions the FRIA can ``rely on previously conducted fundamental rights impact assessments or existing impact assessments carried out by provider'', which we interpret as the case where previous FRIA are used as inputs. Therefore, we suggest reusing \texttt{dpv:hasData} to indicate when existing FRIA act as inputs to the current FRIA process.

Similarly, Article 27-4 refers to the FRIA `complementing a DPIA' where the DPIA covers some obligations related to the FRIA. As before, we also interpret this case as providing for a DPIA to be reused within the FRIA process, which can be expressed by using the relevant DPV concepts to represent DPIA and associating it with a FRIA through the relation \texttt{dpv:hasData}.

In the above, we have modelled our concepts based on the necessity to document the information as required within the AI Act. An alternative method to document these obligations is through the use of the PROV-O ontology \cite{lebo2013prov} where each step is an activity with specific inputs and outputs, and where the provenance of activities and input/output artefacts is to be maintained as logs.

\subsection{Concepts to represent outcomes of FRIA}
AI Act's Article 27-1 states the FRIA's objective is to produce an ``assessment of the impact on fundamental rights that the use of (AI) system may produce'', which in the simplest interpretation implies a boolean categorisation as to whether there is or isn't an impact on fundamental rights. We therefore represent the process of determining the outcome of a FRIA process as the concept \texttt{fria:FRIAOutcome}, which extends the existing \texttt{eu-aiact:FRIA} concept and can be associated with a FRIA using the relation \texttt{dpv:hasAssessment}. And to represent the outcomes, we model these as statuses through the concept \texttt{fria:FRIAOutcomeStatus} which can be associated by using the relation \texttt{dpv:hasStatus}.

We also create instances to represent the specific outcomes possible as per Article 27: 
\setlist{nolistsep}
\begin{enumerate}
  \item \texttt{fria:FRIAOutcomeUnacceptableRisk}: FRIA outcome status indicating that there is an unacceptable risk to fundamental rights, implying the AI system should not be used.
  \item \texttt{fria:FRIAOutcomeHighResidualRisk}: FRIA outcome status indicating high residual risk to fundamental rights which are not acceptable for continuation.
  \item \texttt{fria:FRIAOutcomeRisksAcceptable}: FRIA outcome status indicating residual risks to fundamental rights remain and are acceptable for continuation.
  \item \texttt{fria:FRIAOutcomeRisksMitigated}: FRIA outcome status indicating (all) risks to fundamental rights have been mitigated and it is safe for continuation.
\end{enumerate}

As part of the FRIA procedure and the outcome process, it is essential to identify the relevant fundamental rights which might be impacted. Therefore, we reuse the DPV's extension modelling the EU Charter of Fundamental Rights and Freedoms where each article within the charter is represented as an instance of \texttt{dpv:Right}. To represent which right is impacted, we reuse the concept \texttt{risk:ImpactToRights} along with the relevant instance of fundamental right, and associate it by using the relation \texttt{dpv:hasImpact}.

To enable the granular investigation of impact on rights as required in the FRIA process, we identify the need to to create impact concepts for each right in a manner that allows directly stating that right has been impacted e.g. \textit{Impact on Right of Non-Discrimination}. We also argue that it would be useful to further represent such impacts at an even more granular level by creating concepts representing impacts on specific \textit{requirements} within the right e.g. to state there has been an impact on this right due to discrimination based on a specific category such as sex, race, gender, etc. as mentioned in Article 21 of the Charter. We propose such concepts be added to the DPV's extension modelling fundamental rights so that they can be readily used with the rest of DPV's risk and impact assessment concepts.

\subsection{Concepts to represent notification of FRIA}
AI Act Article 27-3 states that upon completion of a FRIA, a deployer ``shall notify the market surveillance authority of its results''. We represent this step as the concept \texttt{fria:FRIANotificationAssessment}, which extends the existing \texttt{eu-aiact:FRIA} concept and can be associated with a FRIA using the relation \texttt{dpv:hasAssessment}. This steps requires an assessment of whether the notification is required to be sent, or if there is an exception as Article 27-3 also states ``In the case referred to in Article 46(1), deployers may be exempt from that obligation to notify''. 

To represent whether a notification is needed and has been communicated, or is exempt, we create the concept \texttt{fria:FRIANotificationStatus} which extends \texttt{dpv:Status} and its instances:
\setlist{nolistsep}
\begin{enumerate}
  \item \texttt{fria:FRIANotificationNeeded} for when a notification has been identified as being needed, and requires further assessment for whether it is required to be sent or is exempt; 
  \item \texttt{fria:FRIANotificationNotSent} for when a FRIA notification is identified as being required but it not sent yet;
  \item \texttt{fria:FRIANotificationSent} for when the notification is sent; and 
  \item \texttt{fria:FRIANotificationExempt} for which the obligation to notify is exempt as per Article 46-1. As each market surveillance authority will have the ability to create exemptions, we also note the possibility to expand this concept for different jurisdictions based on the DPV's modular legal framework.
\end{enumerate}

DPV contains several notices for obligations under GDPR such as for privacy notice, data breach reporting, rights exercise notices - and also provide guidance on modelling the information and metadata involved. We therefore reuse this method of using `notices' to indicate communication of information between entities, and represent information in a FRIA notification the concept \texttt{fria:FRIANotice} by extending \texttt{dpv:Notice}, which can be associated using \texttt{dpv:hasNotice}.

\subsection{FRIA Questionnaire and Automated Tool}
AI Act Article 27-5 states the existence of a questionnaire based on a template that the deployers can use to complete the FRIA obligations, such as in Article 27-3 for communicating the FRIA to market surveillance authorities by submitting the filled out questionnaire template. 
The AI Act Article 27-5 further states that the questionnaire, ``including through an automated tool'', is intended ``to facilitate deployers in complying with their obligations under this Article'' - which means that the FRIA questionnaire and documentation in some part can be based on automated tools.

To represent these processes, we find two interpretations based on the word `template' having two meanings. First interpretation consists of three artefacts - a template questionnaire used to create a questionnaire, which is then used by deployers to create a filled out questionnaire. The second interpretation consists of two artefacts - a questionnaire is used by the deployer to create a filled out questionnaire. We follow it is the second interpretation, and represent it through the concepts \textit{fria:FRIAQuestionnaire} to represent the template questionnaire that is given to be filled out by the deployer, and extend it as \textit{aiact:FRIACompletedQuestionnaire} to represent the filled out questionnaire which the deployer can send in their notice to the market surveillance authority.

To represent the involvement of tools as per Article 27-5, we create the conecpt \textit{aiact:FRIATool} which extends \texttt{dpv:Technology}. This enables the reuse of DPV TECH extension concepts regarding the modality (e.g. service, product), stakeholders (e.g. developer, user), and other relevant data to be modelled using existing concepts. The tool can be represented as being used in relevant steps of the FRIA by associating it with the \texttt{dpv:isImplementedUsingTechnology} relation. 

We do not think it is necessary to explicitly define the tool as being automated given the purpose of the ontology is to create information systems which by definition use automation in some form. That being said, automated here can be interpreted to have different meanings within the context of ``to facilitate deployers in complying with their obligations under this Article in a simplified manner''. The tool can be used to determine necessity, to assist in collecting and organising input information, to determine - manually or through reasoning - whether there is an impact on rights, and to support notification to authority \textit{aiact:FRIANotificationProcedure}.






\section{Conclusion \& Future Work}
Our work represents the first ontology for modelling the Fundamental Rights Impact Assessments (FRIA) under the AI Act in terms of the information involved as well as the procedure for conducting FRIA itself. As we utilised well-established standards in the legal domain - namely the semantic web standards - our ontology enables the use of machine-readable information that is interoperable, extensible, and well-structured by default. Through this, we enable the creation of automated tools to assist with the FRIA processes that can use our ontology to structure and use information in a consistent manner across use-cases. By being a semantic web ontology, our approach permits using existing standards for information retrieval such as SPARQL, and for SHACL for validation. It also facilitates use of logical semantic reasoning to infer additional information - such as specific risks and impacts, and to ensure completeness and correctness of information.

We developed our ontology by utilising and extending the Data Privacy Vocabulary (DPV), which is a state of the art resource that is continuously developed, and provides a modular approach to representing legal concepts across jurisdictions. This enables the development of practical tools that not only address the FRIA, but are also compatible with the existing and potential uses of DPV in legal processes associated with compliance, documentation, and communication. Our future work therefore consists of integrating our proposed concepts in DPV by participating within the W3C Data Privacy Vocabularies and Controls Community Group (DPVCG), developing a prototype FRIA questionnaire and automated tool that uses and maintains information based on our ontology and DPV, and performing experiments to understand its utility for different stakeholders such as organisations, auditors, market surveillance authorities, and the AI Office.

\begin{acknowledgments}
This work was funded by the ADAPT SFI Centre for Digital Media Technology, which is funded by Science Foundation Ireland through the SFI Research Centres Programme and is co-funded under the European Regional Development Fund (ERDF) through Grant\#13/RC/2106\_P2. Harshvardhan J. Pandit is the current chair of the W3C Data Privacy Vocabularies and Controls Community Group (DPVCG) and the editor/maintainer of Data Privacy Vocabulary (DPV).
\end{acknowledgments}

\bibliography{references}
\end{document}